\documentclass[sigconf, nonacm, 10pt]{acmart}

\AtBeginDocument{%
  \providecommand\BibTeX{{%
    \normalfont B\kern-0.5em{\scshape i\kern-0.25em b}\kern-0.8em\TeX}}}
\setcopyright{none}
\usepackage{enumitem}

\newcommand\system{Proteus}

\usepackage[nointegrals]{wasysym}
\usepackage{multirow,booktabs,tabu,float}
\usepackage{graphicx}
\usepackage{subcaption}

\begin{document}

\title{Time-bound Contextual Bio-ID Generation for Minimalist Wearables}

\author{Adiba Orzikulova}
\affiliation{
  \institution{KAIST, Republic of Korea}
  \country{}
}
\email{adiorz@kaist.ac.kr}

\author{Diana A. Vasile}
\affiliation{
  \institution{Nokia Bell Labs, UK}
  \country{}
}
\email{diana-alexandra.vasile@nokia-bell-labs.com}

\author{Fahim Kawsar}
\affiliation{
  \institution{Nokia Bell Labs, UK}
  \country{}
}
\email{fahim.kawsar@nokia-bell-labs.com}

\author{Chulhong Min}
\affiliation{
  \institution{Nokia Bell Labs, UK}
  \country{}
}
\email{chulhong.min@nokia-bell-labs.com}

\begin{abstract}
  As wearable devices become increasingly miniaturized and powerful, a new opportunity arises for instant and dynamic device-to-device collaboration and human-to-device interaction. However, this progress presents a unique challenge: these minimalist wearables lack inherent mechanisms for real-time authentication, posing significant risks to data privacy and overall security. To address this, we introduce \system{} that realizes an innovative concept of \emph{time-bound contextual bio-IDs}, which are generated from on-device sensor data and embedded into a common latent space. These bio-IDs act as a \emph{time-bound unique user} identifier that can be used to identify the wearer in a certain context. \system{} enables dynamic and contextual device collaboration as well as robust human-to-device interaction. Our evaluations demonstrate the effectiveness of our method, particularly in the context of minimalist wearables.
\end{abstract}




\maketitle
\section{Introduction}

Wearable technology has emerged as an integral part of modern society, with a vast array of applications ranging from health monitoring to activity tracking and beyond. As these devices shed screen-based interactions and integrate ML through tiny AI accelerators~\cite{moss2022ultra}, they are becoming increasingly miniaturized, while still remaining powerful. Examples include smart earbuds, smart rings, and fitness bands—what we term ``minimalist wearables''. These tiny form factors make it likely that we will soon find ourselves surrounded by an ever-growing number of such devices.

This trend opens new opportunities in device-to-device collaboration and human-to-device interaction. For instance, dynamic and contextual device collaboration can lead to enriched and seamless context monitoring in a more accurate and energy-efficient manner~\cite{min2019closer}. Additionally, the nature and multiplicity of these wearables make them ideal for shared use, such as listening to music together through a single pair of earbuds or sharing smart wristbands in a factory setting. Despite such benefits, minimalist wearables present a unique challenge: they lack inherent mechanisms for instant and dynamic authentication for users and devices. Typically, during the initial Bluetooth pairing with a user's device, wearables are associated with a user's account and rely on this static link for all subsequent interactions and data management. Once worn, these wearables operate without additional authentication steps. However, this lack of real-time verification poses significant risks, such as unauthorized usage or misinterpretation of data, thereby highlighting a substantial vulnerability and compromising both data privacy and the overall security~\cite{bianchi2016wearable, mendoza2018assessment}.

To address this risk, we propose an innovative concept, \emph{time-bound contextual bio-ID}, which can support dynamic and contextual device collaboration and interaction on the fly. These bio-IDs are representations of sensor data embedded into a common latent space, ideally universal for an individual irrespective of device placement, but distinct for different users or at different times. We develop \system{}, a framework that generates time-bound contextual bio-IDs from Inertial Measurement Unit (IMU) and Photoplethysmography (PPG) sensors. The key idea is to extract a common latent space of sensor data by leveraging contrastive learning~\cite{chen2020simple}. More specifically, in our framework, sensor data from multiple wearables worn by the same user at the same time serve as positive pairs for contrastive learning, encouraging similar embeddings, while data from different users or different time points act as negative pairs, pushing for disparate embeddings. This time-bound contextual bio-ID concept ensures we have accurate embedded representations to enable robust human-to-device and device-to-device matching even in minimalist wearables.

\section{Motivation and Challenges}

We discuss the driving motivation behind generating \emph{time-bound contextual Bio-IDs} and illustrate their usefulness through a discussion of use cases. Lastly, we review the main challenges to create the Bio-IDs.

\textbf{Why Time-Bound?} 
As our digital ecosystem continues to grow increasingly interconnected, the urgency for secure, efficient, and adaptable authentication methods intensifies. Traditional authentications, like passwords or biometrics, offer high security but are static in nature, lacking the ability to adjust dynamically based on real-time context or instant device-to-device collaboration. To address this, we propose an innovative concept of \emph{time-bound contextual bio-IDs}. This approach not only provides robust security through the fusion of vital signs but also incorporates a layer of temporal flexibility. This time-sensitive aspect enables dynamic access control, streamlines multi-device authentication, and allows for personalized, context-aware multi-device collaboration. 

\textbf{Why Vital Signs?} 
They are an intuitive choice for the time-bound contextual bio-IDs, particularly when utilizing the functionalities of current wearable technology. These wearables are commonly outfitted with a diverse range of sensors that are tailored to monitor human physiology and behavior in real-time. Such devices record key metrics including heart rate, heart rate variability, oxygen saturation, respiration rate, and blood pressure~\cite{ferlini2021ear}. These metrics are inherently universal for an individual regardless of the device placement, but unique according to individual's dynamic and contextual factors~\cite{chang2020systematic, cahoon2023continuous}.

These vital signs are already monitored for health and wellness purposes, e.g., using PPG, and are also affected by physical activity, which can be measured through existing IMU sensors. By using these common sensors, we can generate time-bound contextual bio-IDs without the need for extra hardware or complicated user interactions.

\subsection{Use cases}~\label{subsec:use_cases}
\textbf{Instantaneous device association.} The proposed system allows for immediate and secure device association using bio-IDs. For example, upon purchasing a new smart ring, users can easily associate it with their smartphone. By placing the ring-wearing finger against the rear camera of the smartphone, vital signs such as heart rate and heart rate variability can be captured both on a smartphone~\cite{ayesha2021heart} and a smart ring~\cite{zhou2023one}. These vital signs are then used to generate bio-IDs that can be compared with each other, facilitating immediate and secure device pairing without manual configuration.

\textbf{Seamless authentication across wearables.} Current minimalist wearables lack continuous authentication capabilities, posing potential security risks, especially in shared-use scenarios. For example, today's smart earbuds are not capable of discerning whether they are being used by the authenticated owner or someone else. \system{} provides a solution by continuously comparing vital signs from left and right earbuds and verifying the user. This allows the system to filter out sensitive information to dynamically adapt to different users, e.g. notifying an incoming message rather than reading out its contents.

\textbf{Dynamic access control for contextual device collaboration.} Typically, once wearables mutually authenticate, they operate under the assumption that a secure connection is maintained indefinitely. Our proposed system challenges this static approach by incorporating dynamic, time-sensitive access control based on real-time bio-ID verification. 
If a smartwatch app aims to collaborate with earbuds for body gesture recognition while the user is sitting, time-bound access to the earbuds' data can be granted accordingly, thereby preventing the app from accessing the earbuds at an unwanted time.

\subsection{Challenges}\label{subsec:challenges}

\begin{figure}[t]
  \centering
    \includegraphics[width=\linewidth]{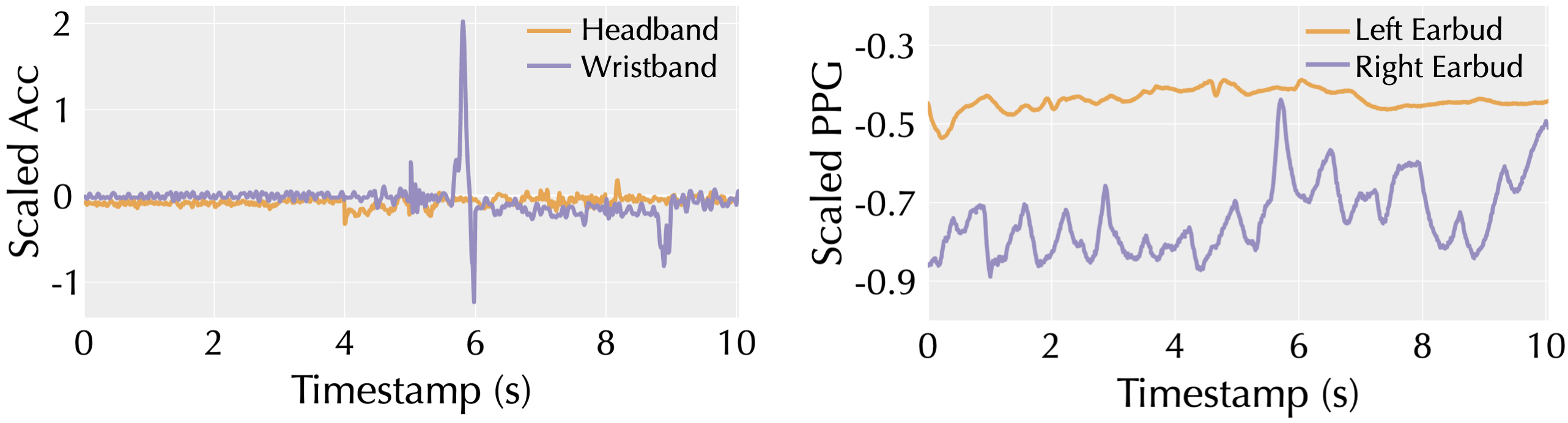}
    \vspace{-0.2in}
    \caption{Accelerometer from headband and wristband (left) and PPG from left and right earbuds (right).}
    \label{fig:motivating_example}
    \vspace{-0.2in}
\end{figure}

\begin{itemize}[leftmargin=*]
    \item \textbf{Device and sensor heterogeneity}. The first challenge lies in the heterogeneity of device architecture and sensor placement. This results in the same physiological and behavioral contexts being captured differently across various devices (see Figure~\ref{fig:motivating_example}), complicating the process of fusing sensor data to generate a reliable and contextual bio-ID. 
    \item \textbf{Scalability concerns}. As the number of wearable devices in the ecosystem increases, the task of managing unique bio-IDs for each sensor, device, user, and activity becomes increasingly complex and computationally intensive, posing significant scalability challenges.
    \item \textbf{Immediate usability requirement}. The system must have the capability to generate a bio-ID immediately upon device deployment, and this needs to occur without requiring any additional intervention from the user, adding a layer of complexity to system design and implementation.
\end{itemize}

\section{\system{}}

A straightforward way to generate bio-IDs would be to extract and combine device/sensor-specific features or absolute values of various vital signs. 
However, sensor readings or features from these devices exhibit variability due to heterogeneity in device location, hardware specifications, and software characteristics. Conversely, establishing a unique bio-ID for every potential device combination would encounter significant scalability challenges~(\S\ref{subsec:challenges}).
We introduce the \emph{universal embedding bio-ID generation} approach: generating embeddings from live sensor data such that the bio-ID appears similar across different devices worn by the same user at the same time. This approach allows us to create a unique, time-sensitive bio-ID by focusing on common embeddings rather than specific features or absolute values of vital signs. The benefits of this method are threefold: (1) It addresses device and sensor heterogeneity by creating normalized, embedded representations that are common across various devices.
(2) It simplifies the system architecture and alleviates the need to manage numerous model combinations, effectively addressing scalability issues.
(3) It allows for an immediate bio-ID generation as soon as the device is deployed, without requiring additional user-side training. 

There are various methods for extracting shared embeddings from diverse signals, including 
autoencoders~\cite{makhzani2015adversarial, kingma2019introduction}, Siamese networks~\cite{chicco2021siamese}, and contrastive learning~\cite{chen2020simple, he2020momentum, oord2018representation}.
In this paper, we have chosen contrastive learning for several compelling reasons: (a)~it has recently demonstrated impressive capabilities in extracting robust representations, (b)~it possesses the ability to capture intricate nonlinear relationships, and (c)~we can naturally adapt it to multi-device environments.
Contrastive learning is primarily employed in self-supervised machine learning to learn representations by contrasting positive and negative examples
~\cite{chen2020simple, he2020momentum}
. In a typical setting, similar data points are pulled closer in the embedding space, while dissimilar ones are pushed apart. For our application, this framework naturally aligns with the availability of sensor data from multiple wearable devices used by a single user. Such data can be treated as \textit{positive} pairs, encouraging the model to generate similar embeddings for them. Data from different users or from the same user at different time points can be considered as \textit{negative} pairs, driving the model to produce disparate embeddings.

\subsection{System Operation}

\system{} operates as follows: Initially, a global bio-ID model is generated on the server side, trained using a global dataset that accounts for various sensor combinations. Upon deploying devices to new users, appropriate models are downloaded based on the sensors available, and bio-IDs are produced by running these models on real-time sensor data. The model can be personalized and updated with individual user data, which we leave as future work.

\textbf{Adaptive Model Selection.} To generate the bio-ID model, we face a design trade-off between model complexity and real-time performance. One strategy would be to train separate models for each sensor type. While this offers simplicity in model design and flexibility in deployment, it limits the model's ability to leverage inter-sensor relationships and correlations, thus potentially sacrificing performance. Another approach would be to construct a monolithic model that ingests data from an entire set of possible sensors on wearables. However, this is not practical for real-life applications due to the variations in available sensors across different devices. For example, filling zero values for unavailable sensors would lead to potentially lower runtime performance.
To address these challenges, we dynamically select the most appropriate embedding models based on the overlapping sensors available at runtime. This allows for maximum flexibility and performance optimization but comes at the cost of a training overhead. We consider this overhead acceptable as the computational burden falls on the server side and needs to be carried out only once, prior to deployment.

\begin{figure}[t]
  \centering
  \begin{minipage}[b]{0.45\textwidth}
  \centering
      \includegraphics[width=\textwidth]{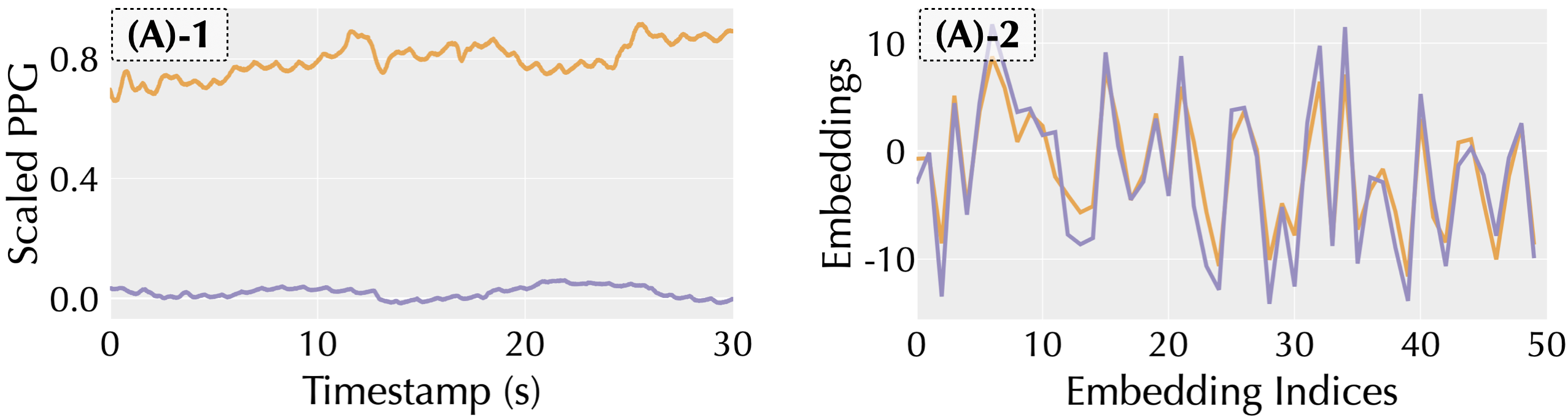}
  \end{minipage}
  \hfill
    \begin{minipage}[b]{0.45\textwidth}
  \centering
      \includegraphics[width=\textwidth]{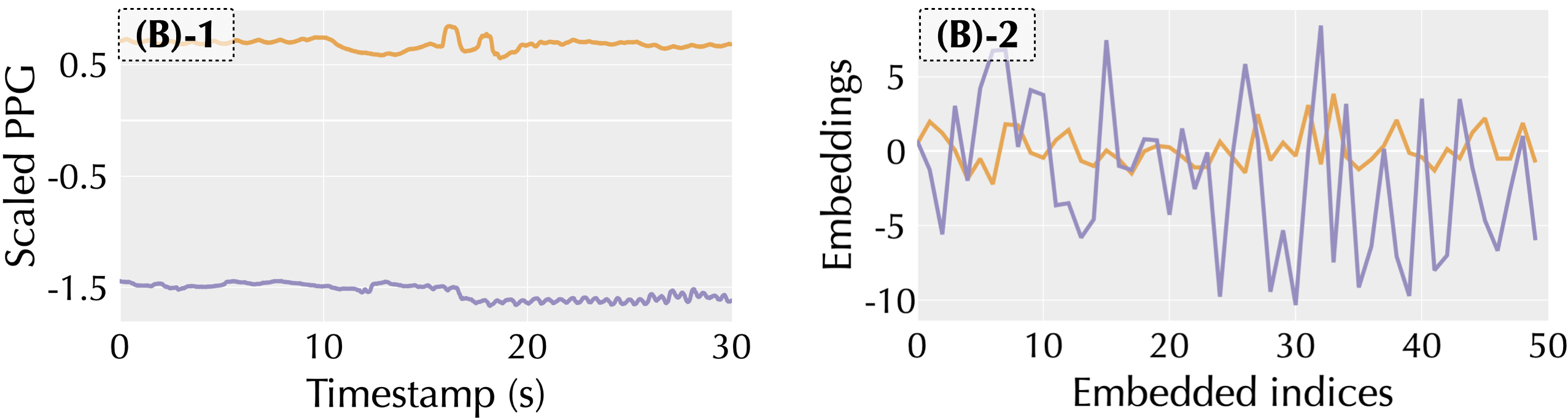}
  \end{minipage}
\hfill
    \begin{minipage}[b]{0.445\textwidth}
  \centering
      \includegraphics[width=\textwidth]{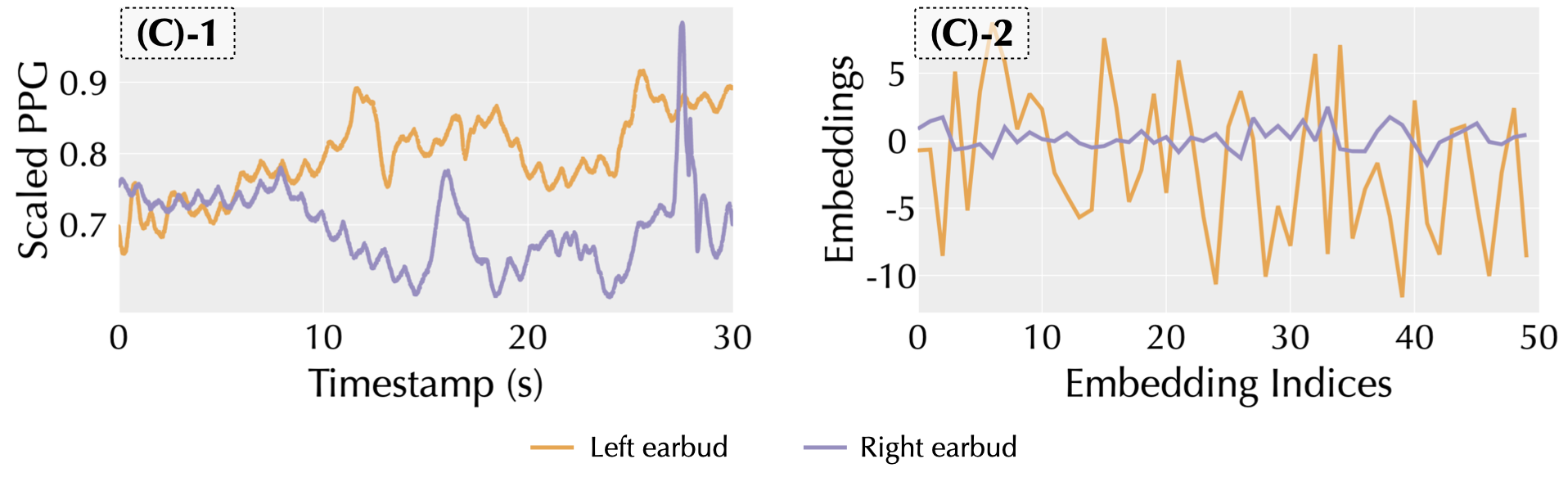}
  \end{minipage}
  \caption{Example of PPG (left) and embeddings (right) on two earbuds. (A) Same user, time-aligned; (B) Different users; (C) Same user, non time-aligned.}
\label{fig:embeddings_qualitative}
\vspace{-0.15in}
\end{figure}

\textbf{Bio-ID Model Training.} Recently, contrastive learning-based techniques emerged to learn common embedded representations from sensor data such as Contrastive Predictive Coding~\cite{haresamudram2021contrastive}, SimSiam~\cite{chen2021exploring}, 
and SimCLR~\cite{chen2020simple, tang2020exploring}. Our approach builds upon SimCLR~\cite{chen2020simple} due to its capacity to accommodate varying batch sizes, thereby reducing memory consumption. SimCLR demonstrates promising outcomes when applied to smaller datasets~\cite{tang2020exploring}, unlike the vast volumes of data available in vision or audio domains. 
We adapt original SimCLR as follows: we designate time-aligned sensor data of the same user from multiple wearable devices as \textit{positive pairs}, while treating data from other users and different time instances as \textit{negative pairs}. When the number of devices associated with a given set of available sensors is more than two, we enhance the neural network's generalization capability by randomly selecting two distinct devices in each batch training. The embedded representations are structured in the form of 1D arrays and are further processed for matching purpose. Even when the raw data does not show correlation, the resulting embeddings are expected to be aligned when the devices are on the same body at the same time and not aligned when the data was captured from the same body but at different times, or on different bodies (see Fig.~\ref{fig:embeddings_qualitative}).

Our model's network architecture is inspired by recent work~\cite{tang2020exploring}. The base network consists of three convolutional blocks and one max-pooling layer. Each convolutional block includes a 1D Convolutional layer, ReLU activation, BatchNorm, and Dropout. A projection head with three fully connected layers along with ReLU activation layers is attached to the base network. To extract representations for time-bound contextual bio-IDs in the latent space, we optimize our objective with SGD, starting with a learning rate of 0.1 and progressively reducing it via Cosine Decay. Our model undergoes training for a device- or sensor-specific epochs until it converges.

\textbf{Bio-ID Matching.} The next critical step is to perform efficient and accurate Bio-ID matching between different devices (see the use cases in \S\ref{subsec:use_cases}). 
We employ a set of fully-connected layers that are designed to take two embeddings as input and output a binary label, indicating whether the embeddings match (``matched'') or not (``unmatched'').
We train these fully-connected layers using labeled data, ensuring that the model generalizes well to real-world scenarios. This architecture is particularly beneficial for \system{}, as it allows us to extend the matching process to an ensemble learning approach for global decision-making when more than two devices are involved.

\section{Evaluation}

We first assess the bio-ID generation process, focusing on how similar bio-IDs are created to each other even with device placement heterogeneity. 
Then, we evaluate the performance of bio-ID matching, considering different combinations of sensors and devices, and different user activities. 

\subsection{Experimental Setup}

\paragraph{Dataset}
The experiments are run using the FatigueSet dataset~\cite{kalanadhabhatta2021fatigueset}, which encompasses a diverse range of multi-device (earbuds, wristband, headband) and multi-sensor (IMU, PPG) data collected from twelve participants performing various physical and mental activities. The PPG data obtained from earbuds was sampled at a rate of 100 Hz, while the IMU data was sampled at rates of 100 Hz, 52 Hz, and 32 Hz from earbuds, headband, and wristband, respectively. To ensure consistency, we re-sampled all data at 100 Hz, applied standard scaling to normalize it, and then divided the normalized data into segments (without overlap) with window sizes of 20 seconds (IMU only) and 30 seconds (PPG only, IMU/PPG). 

\paragraph{Metrics} We quantitatively assess the performance of \system{} by using the True Positive Rate (TPR), False Positive Rate (FPR), and False Negative Rate (FNR) to measure the effectiveness and reliability of bio-ID generation and matching in different scenarios. Here, TPR measures the rate at which genuine users are successfully authenticated by our system, FPR the rate of incorrect authentication (for imposters), and FNR the rate of rejection for genuine users (resulting in decreased usability). An ideal bio-ID generation system would thus maximise TPR and minimise FPR and FNR.

\subsection{Bio-ID Generation}

\begin{figure}[t]
  \centering
    \includegraphics[width=\linewidth]{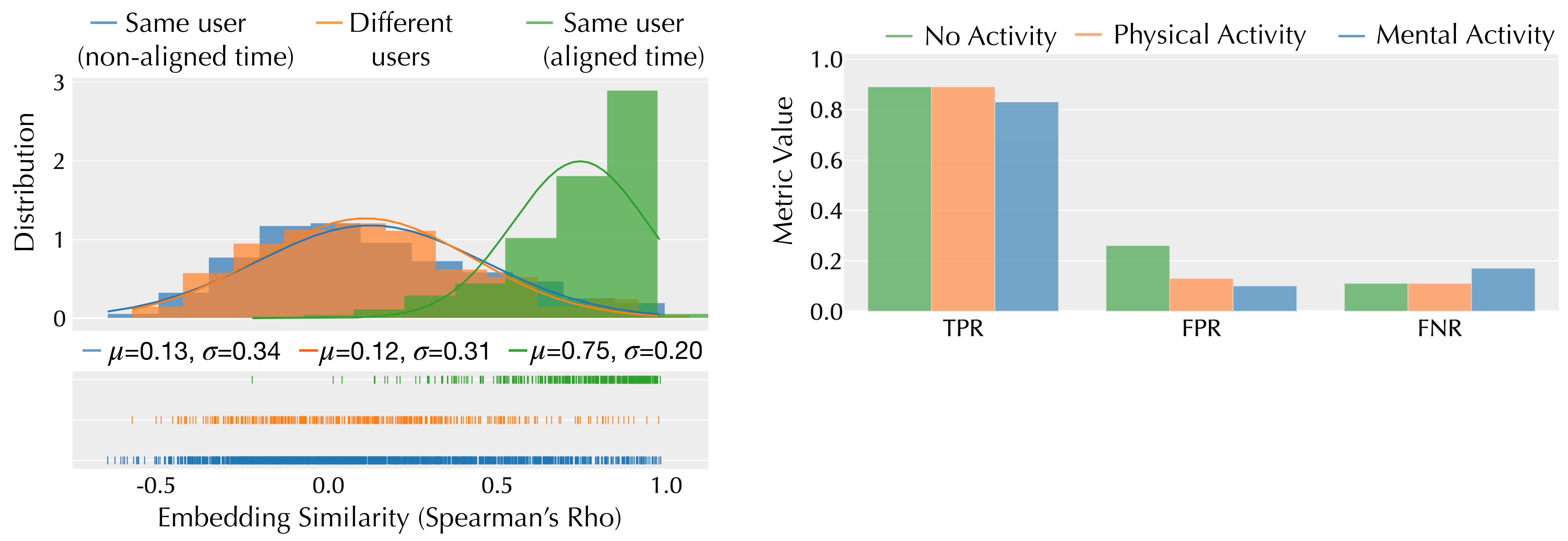}
    \caption{Bio-ID Generation (Left) and Matching (Right)}
    \label{fig:generation_quality_distribution}
    \vspace{-0.2in}
\end{figure}

We conducted a thorough analysis of the bio-ID generation process to evaluate its performance in terms of uniqueness, robustness, and temporal stability. This analysis aimed to ensure that the generated bio-IDs accurately represented individual wearer patterns and remained consistent to support the bio-ID matching process over time.

For this purpose, we define the following experiment on the FatigueSet user's data --- we split the dataset into three different groups: group~A represents the sensor readings from the same user's two devices at different times; 
group~B represents the data coming from two different users' devices; and group~C represents the sensor readings from the same user's two different devices at the same time. Fig.~\ref{fig:generation_quality_distribution} (left) shows the distribution of the embedding similarity, calculated using Spearman's rank correlation coefficient (also known as Spearman's $\rho$ )~\cite{zar2014spearman}. We chose to calculate Spearman's $\rho$, as a non-parametric measure that does not assume any specific underlying data distribution. Values close to 1 indicate a strong positive similarity, while values near -1 suggest strong negative dissimilarity, and values around 0 denote no discernible similarity.

The experimental results for Group C, which includes embeddings of the same user from two different devices acquired simultaneously, show that the data is centered around the right tail ($\mu$=0.75, $\sigma$=0.20). This compellingly demonstrates that our method maps raw data from two distinct devices and locations of the same user at the same time to a similar representation space. In contrast, we can see that for Group A ($\mu$=0.13, $\sigma$=0.34) and Group B ($\mu$=0.12, $\sigma$=0.31), the data is clustered around 0. This implies that the generated embeddings are distant from each other in the latent space.

\subsection{Bio-ID Matching}

We further validate the effectiveness of \system{} with bio-ID matching. To examine the overall matching performance, we split the dataset into two groups: group~1 represents the IMU and PPG data from left and right earbuds, and group~2 represents the IMU data from four devices (left and right earbuds, headband, wristband). Our results provide an overall promising start. First, when two sensors are used together in group~1, \system{} shows a consistent correct matching with high TPR of 84\% but with low FPR of 9\% and FNR of 16\%. We further investigate the matching performance only when the IMU data is used in group~2. Despite the influence placement has on IMU sensors, we still see a reasonable correct matching over in over 70\% of cases (see first row of each sensor selection in Table~\ref{table:results}). 

\footnotesize
\begin{table}[t]
    \centering
    \begin{tabu}{p{2cm} {c}*{8}{p{0.2cm}} p{1cm}}   
      Sensor Selection & & \multicolumn{4}{c}{Device Placement} & 
             \multicolumn{5}{c}{Results} \\
	\midrule
         
    \multirow{3}{*}{} 
    & 
    & \rotatebox{45}{\textbf{L. Ear}} 
	& \rotatebox{45}{\textbf{R. Ear}}
	& \rotatebox{45}{\textbf{Head}}
	& \rotatebox{45}{\textbf{Wrist}}
    & \rotatebox{45}{\textbf{TPR}}
    &
    & \rotatebox{45}{\textbf{FPR}}
    &
    & \rotatebox{45}{\textbf{FNR}} \\
	\midrule 

   \textbf{Accelerometer}
    & &\LEFTcircle &\LEFTcircle &\LEFTcircle &\LEFTcircle &0.71 &&0.22 &&0.29   \\
    & & -- &-- &\newmoon &\newmoon     &0.63 &&0.24 &&0.37   \\ 
    & & \newmoon &-- &\newmoon & --     &0.79 &&0.26 &&0.21   \\  
    & & \newmoon &\newmoon &--  & --    &0.81 &&0.26 &&0.19   \\    
    & & \newmoon  &--  & -- &\newmoon   &0.60 &&0.24 &&0.40   \\ 
    & &--  & \newmoon & \newmoon & --   &0.82 &&0.22 &&0.18   \\ 
    & &--  & \newmoon & -- & \newmoon   &0.60 &&0.23 &&0.40   \\ 
	\midrule
    
    \textbf{Gyroscope}
    & &\LEFTcircle &\LEFTcircle &\LEFTcircle &--  &0.72 &&0.15 &&0.28   \\
    & & \newmoon &-- &\newmoon & --     &0.72 &&0.21 &&0.28   \\  
    & & \newmoon &\newmoon &--  & --    &0.80 &&0.17 &&0.20   \\     
    & &--  & \newmoon & \newmoon & --   &0.64 &&0.16 &&0.36   \\ 
    \midrule

    \textbf{Acc \& Gyro}
    & &\LEFTcircle &\LEFTcircle &\LEFTcircle & -- &0.76 &&0.22 &&0.24   \\
    & & \newmoon &-- &\newmoon & --     &0.77 &&0.22 &&0.23   \\  
    & & \newmoon &\newmoon &--  & --    &0.81 &&0.24 &&0.19   \\     
    & &--  & \newmoon & \newmoon & --   &0.69 &&0.28 &&0.31   \\
   \bottomrule
    \end{tabu}
  \caption{Results of experiments\\ \newmoon~=~selected, \LEFTcircle~=~randomised selection, --~=~not selected }
  \label{table:results}
  \vspace{-0.8cm}
  \end{table}
\normalsize

\subsubsection{Effect of Physical and Mental Activities}

First, the FatigueSet dataset splits the data into three distinct phases: no activity, physical activity, and mental activity. \textit{No activity} is an initial phase of the data capture where users relax (still and not performing any intensive mental activity either). We run our Bio-ID generation and matching for the sensor data in each of these different activity types. Fig.~\ref{fig:generation_quality_distribution} (right) shows the bio-ID matching performance for the different activities. All available sensors (IMU and PPG) were used in this experiment. We observe consistent TPR greater than 80\% -- it ranges between 83\% to 89\% across each activity type. 

\subsubsection{Effect of Device Placement}

We look in detail at the effect of device placement (left and right ears, head or wrist) and sensor combination (accelerometer and gyroscope). The results are shown in Table~\ref{table:results}. Accelerometer was available in all four devices, while gyroscope was not available in wristband, so the combination of accelerometer and gyroscope was also used in at most three devices. When we randomly select two devices out of four or three available devices, the TPR is always higher than 70\%, while offering FPR and FNR rates lower than 22\% and 29\%, respectively. Across various sensor configurations, it becomes apparent that optimal matching is achieved when the selected devices are located in close proximity and exhibit similar movement patterns (e.g., left and right earbuds). In such scenarios, the TPR exceeds 80\%, with FPR and FNR both under 26\% and 20\%, respectively. Conversely, when the selected devices are positioned far apart (e.g., the right ear and wrist), the matching performance tends to decline. Additionally, combining accelerometer and gyroscope data tends to yield superior results.

\section{Related Work}

Existing passwordless device authentication relies on specialized hardware like fingerprint readers, retina scanners, or custom biometric modules. These systems typically come with higher costs, limited device compatibility, and a focus on human-to-device authentication rather than device-to-device interaction.
There have been active research efforts to realise authentication with wearable sensors such as motion sensors, PPG, ECG, and EMG by capturing unique behavioral or physiological patterns~\cite{blasco2016survey}. For instance, user verification may involve gait analysis with motion sensors on smartwatches~\cite{xu2017gait} and earbuds~\cite{ferlini2021eargate}, or heart rate analysis with PPG~\cite{vhaduri2017wearable} and ECGs~\cite{chun2016ecg}.
Despite the advancements in wearable-based authentication, our approach stands out in two key respects. Firstly, it is designed to be time-bound and contextual, adapting to dynamic shifts in a user's behavior and environment. Secondly, our method can be immediately deployed without any user interaction, eliminating the need for user intervention during the enrolment process.

\section{Limitations and future work}

\textbf{Dataset}: We observe that the dataset consists of a relatively small pool of 12 participants, which raises questions about the generalizability of our findings. Additionally, the sensor types are confined to IMU and PPG, excluding other potential sensing modalities that could provide more nuanced insights. Furthermore, the devices were deployed on only four specific body positions, limiting the breadth of scenarios. While the outcomes are indeed promising, these dataset limitations emphasize the need for future work involving more extensive and diversified datasets.

\textbf{Heterogeneous sensor sets and application requirements}: The current study focuses on scenarios where two devices feature the same set of sensors, thereby simplifying the embedding generation and matching process. However, the complexity of these operations increases substantially when dealing with multiple wearables equipped with varying sensor types. 
Additionally, different apps may have varying requirements for authentication; for example, casual apps might prioritize energy efficiency, while mission-critical apps may demand high accuracy. These aspects necessitate a more dynamic and context-aware approach for bio-ID generation and matching, opening up avenues for future research.

\textbf{Privacy and security concerns:} While our time-bound contextual bio-IDs enable more secure and dynamic device/user authentication, they might still pose privacy and security concerns, especially when embeddings are shared between (malicious) devices. In its current form, the system directly transmits raw embedding data, potentially exposing sensitive biometric information. To address this challenge, we can consider the incorporation of hashing algorithms and encryption into \system{}. By converting these sensitive embeddings into these formats, we could enhance security and privacy measures while maintaining the overall efficacy.

\section{Conclusion}
We use everyday wearable sensors to provide a cost-effective way towards the generation of time-bound contextual bio-ID. Our evaluation highlights the generation accuracy to provide close embeddings when the data comes from devices on the same body at the same time, and versatility across various wearable configurations and activities. This advancement marks a significant step towards enabling seamless minimalist wearable authentication, promising a secure and seamless user-device interaction and device-device collaboration.

\bibliographystyle{ACM-Reference-Format}
\bibliography{mda}

\end{document}